\documentclass[preprint,12pt,amsmath,amssymb,showkeys]{revtex4} 
\usepackage{amsmath}
\usepackage{latexsym}
\usepackage{amssymb}
\usepackage{graphics,epstopdf}
\usepackage{amsmath,amssymb,amsthm}
\theoremstyle{plain}

\usepackage{graphicx}
\usepackage{enumerate}
\usepackage[colorlinks=true, citecolor=blue, urlcolor=blue ]{hyperref}
\usepackage[title]{appendix}
\usepackage{cleveref}

\begin{document}
\title{ Tuning the separability in noncommutative space}

\author{Pinaki Patra}
\email{monk.ju@gmail.com}
\affiliation{Department of Physics, Brahmananda Keshab Chandra College, Kolkata, India-700108}

\date{\today}

\begin{abstract}
We study the Separability of the noncommutative (NC) space coordinate degrees of freedom with the generalized Peres-Horodecki separability criterion (Simon's condition) for a bipartite Gaussian state. Non-symplectic nature of the transformation between the usual commutative space and NC space restricts the use of Simon's condition in  NCS. We transform the NCS system to an equivalent Hamiltonian in commutative space through Bopp shift, which enables the utilization of the separability criterion in NC space. 
For afairly general study, we consider a bilinear Hamiltonian with time-dependent (TD) parameters, along with a TD external interaction, which is linear in field modes. The system is transformed  into canonical form keeping the intrinsic symplectic structure ($Sp(4,\mathbb{R})$) intact.  The solution of the TD-Schr\"{o}dinger equation is obtained with the help of  Lewis-Riesenfeld invariant method (LRIM). Expectation values of the observables (thus the covariance matrix ) are constructed from the states obtained from LRIM.
It turns out that the existence of the NC parameters in the oscillator determines the separability of the states. In particular, for isotropic oscillators, the separability condition for the bipartite Gaussian states depends on  NC parameters. Moreover,  anisotropic parameter values for the oscillator affects the separability. In other words, both the deformation parameters ($\theta,\;\eta$) and parameter values of the oscillator are important  for the separability of bipartite states. Thus tuning the parameter values, one can destroy or recreate the separability of states. With the help of toy models, we have demonstrated TD-NC space parameters effect on separability.
\end{abstract}
 
 \maketitle
\section{Introduction}
Perhaps the most \textquotedblleft genuine\textquotedblright quantum property that a physical system may possess is \textquotedblleft entanglement\textquotedblright, which underlines the intrinsic order of statistical relations between subsystems of a compound
quantum system \cite{entanglement0,entanglement1,schrodinger1}. It occurs in composite systems as a consequence of the quantum superposition principle of states, and of the fact that the Hilbert space that describes a composite quantum system is the tensor product of the Hilbert spaces associated with each subsystem  \cite{entanglement1,schrodinger1,entanglement2,entanglement3}.
In particular, if the entangled subsystems are spatially separated nonlocality properties may arise, showing a very deep departure from classical physics \cite{entanglement4}. In other words, entangled states are inseparable. Initial successful study of necessary and sufficient conditions for separability in the $2 \times 2$ and $ 2 \times 3$ dimensional Hilbert space was done by Peres and Horodecki \cite{separable1,separable2}.  A recent review of the separability criteria in two qubits may also be found in \cite{reviewseparable}.\\
Subsequently, a flood of works in separability conditions for finite-dimensional Hilbert space, more specifically the qubits, have enriched the literature of quantum physics (for a pedagogical survey, see \cite{finiteseparable1}). Simon provided the generalization of the Peres-Horodecki criterion for the separability of a given state of a bipartite canonical continuous system through the variance (correlation) matrix \cite{separable3}. The generalized Peres-Horodecki criterion relies on the basic differences between classical and quantum covariance/noise matrices \cite{separable3,entanglement5}. One of the advantages of Simon's approach is that it is easy to implement in present-day experimental capacity, in particular, the experimental realization of quantum teleportation of coherent states \cite{teleportation1,teleportation2,teleportation3}. 
\\
For a classical probability distribution over a classical $2n$-dimensional phase space, any $2n\times 2n$ real symmetric positive-definite matrix is bonafide, that is, physically realizable, covariance matrix \cite{variance1}. For a quantum system, however, the covariance matrix $(\hat{\mathcal{V}}_c)$ has to satisfy certain additional, namely Robertson-Schr\"{o}dinger uncertainty principle (RSUP) ($	\hat{\mathcal{V}}_c+\frac{i}{2}\hbar\hat{J} \ge 0 $) in the Heisenberg sense, where the symplectic matrix $\hat{J}$ encodes the fundamental commutation relations between co-ordinates $(\hat{q}_k)$ and conjugate momentum $(\hat{p}_k)$ \cite{variance1,rsup1,rsup2,rsup3,rsup4,rsup5,rsup6}. We shall call this background space (position operators commute with each other) as usual commutative space. 
\\
There is a consensus that the space-time structure is deformed in such a manner that the usual notion of commutative space will be ceased at energy scales on the order of $10^{19}$ GeV, i.e., at the energy scale in which the theories of quantum gravity appear to predict departures from classical gravity \cite{qg1,qg2,csseased1,csseased2}. It is almost a Gospel that the fundamental concept of space-time is mostly compatible with quantum theory in NC-space \cite{Witten,Juan}. Therefore, extending ideas of the commutative space quantum mechanics to noncommutative (NC)-space is always an interesting aspect in its own right \cite{ncs1,ncs2,ncs3,ncs4,ncs5,ncs6,ncs7}. For a comparative review of four formulations of noncommutative quantum mechanics may be found in \cite{fourformulation}. In this paper, we study entanglement between coordinate degrees of freedom induced by both the position-position and momentum-momentum noncommutativity parameters, as well as the anisotropy due to the mass and frequency of a quantum mechanical oscillator. In this connection, we would like to mention that the study of noncommutative parameter-induced entanglement for anisotropic oscillators is available in literature \cite{ncinducedentanglement1,ncinducedentanglement2,ncinducedentanglement3,ncinducedentanglement4}.  The present study is distinct from the earlier study in the literature, for its consideration of TD parameters, including the TD-NCS parameters, so that one can envisage the evolution of NC-space by tuning TD-parameters (e.g., TD-magnetic field) for an equivalent experiment in laboratory.  
\\
Since the connection (Darboux transformation) between the commutative space and NC-space is not symplectic, the direct use of generalized RSUP ($\hat{\tilde{\mathcal{V}}}_{nc} + \frac{i}{2} \hbar_e \hat{\tilde{J}} \ge 0$) is not immediate to predict the separability of the bipartite Gaussian state in NC-space \cite{rsupnc1,rsupnc2}. Nonetheless, we can transform the NC-space system into an equivalent system with commutative space operators through Bopp's shift (Darboux transformation) and use the formalism of usual quantum mechanics, including RSUP \cite{rsupnc4,rsupnc5,rsupnc6,rsupnc7}. 
\\
One of the major goals of the present paper is to analyze the effect of parameters in the separability criterion of bipartite state in NC-space. To make our study fairly general, as well as to investigate the experimental possibility by tuning the parameters, we have considered the time-dependent (TD) parameters.
In particular, we have considered a TD-anisotropic harmonic oscillator, placed in an external TD-interaction, which is linear in field modes. From a practical point of view, this is similar to an anisotropic oscillator with a linear external perturbation. For instance, an anisotropic oscillator placed in a TD- field in  $2$-dimensional non-commutative space,  dynamics of a charged particle inside a TD- magnetic field, and Chern-Simon's model- all share a structural similarity in their Hamiltonians \cite{pp1,tdho1,tdho2,tdho3,tdho4,tdho5,tdho6}. The efficacious application of TD-harmonic oscillator (TDHO) in various domains of Physics, makes the study of TDHO a topical one \cite{tdho1,tdho2,tdho3,tdho4,tdho5,tdho6,tdho7}.
\\
The time-dependent system is solved with the help of the Lewis-Riesenfeld (LR) phase-space invariant method \cite{Lewis1,Lewis2,Lewis3,Lewis4,Lewis5,Lewis6,Lewis7,Lewis8,Lewis9,Lewis10}, which states that for a system described by a TD Hamiltonian $H (t )$, a particular
solution of the associated Schr\"{o}dinger equation (SE), is given by the eigenstate ($\vert n,t \rangle $) of a TD-invariant operator $\mathcal{I} (t )$ defined by $\partial_t\mathcal{I}  + (i\hbar)^{-1}[\mathcal{I},\hat{H}]=0$, apart from a TD-phase factor $\exp(i\phi_n(t))$, where $\phi_n(t)=\hbar^{-1}\int_{0}^{t} \langle n,\tau\vert [i\hbar\partial_t -\hat{H}(t)]\vert n,\tau\rangle d\tau.$ The general solution of the SE is
given by the superposition state $\vert \psi(t)\rangle =\sum_{n}c_n\exp(i\phi_n(t))\vert n,t\rangle,$ with $c_n=\langle n\vert \psi(0)\rangle$.
It turns out that the states of the system are displaced number states \cite{displacednumberstate1,displacednumberstate2,displacednumberstate3} along with the TD-phase factor. The exact form of dynamical, as well as geometrical phases, are then determined. The variance (noise/correlation) matrix is then constructed through the expectation values. 
Since, the separability of Gaussian states, may be well characterized through the variance matrix, we have utilized the separability criteria for a bonafide variance matrix.  This leads to some restrictions on the TD- parameters to the bipartite state being separable. We have shown that both the anisotropy of the system and the existence of TD NC parameters are responsible for the entanglement. Thus the separability of the NC-space coordinate degrees of freedom can be destroyed by suitable tuning of the time-dependent parameters. We have analyzed this for toy models. In a first toy model, we show that specific values for the TD deformation parameter $\theta(t)$, the bipartite Gaussian states are entangled. In another toy model, we show that anisotropy through TD-effective mass in NC-space restricts the separability of bipartite states. 
\\
The organization of the paper is the following. Section II corresponds to a general study of the separability criterion in NC-space. The Hamiltonian is put into a canonical form maintaining the $Sp(4,\mathbb{R})$ invariance in section III. The time-dependent system is solved with the help of the Lewis-Riesenfeld invariant method in section IV. The separability criterion through the variance matrix is studied in section V, which is followed by toy models in sections VI and VII.  At last the conclusions of our study are drawn.
\section{ Separability criterion in noncommutative space}
If we define a coordinate vector as
\begin{equation}\label{NCcordefn}
	\tilde{X}= (\hat{\tilde{X}}_1,\hat{\tilde{X}}_2, \hat{\tilde{X}}_3, \hat{\tilde{X}}_4)^T= (\hat{\tilde{x}}_1,\hat{\tilde{p}}_1, \hat{\tilde{x}}_2, \hat{\tilde{p}}_2)^T ,
\end{equation}
then the commutation relations of the observables in NC-space may be represented with a deformed symplectic matrix $\tilde{\Sigma_y}$, given by
\begin{equation}\label{NCspacecommutation}
[	\hat{\tilde{X}}_{\alpha}, \hat{\tilde{X}}_{\beta}]= i\hbar_e \tilde{J}_{\alpha\beta}= - (\tilde{\Sigma}_y)_{\alpha\beta},
\end{equation}
where $\tilde{J}_{\alpha\beta}$ is the $\alpha\beta^{th}$ element of 
\begin{eqnarray}
\hat{\tilde{J}} = \left(\begin{array}{cc}
\hat{J}_2 &  \frac{1}{\hbar_e}\hat{\Pi}_{\theta\eta}\\
 -\frac{1}{\hbar_e}\hat{\Pi}_{\theta\eta} &  \hat{J}_2
\end{array}\right),\; \mbox{with}\; 
\hat{\Pi}_{\theta\eta}= \left(\begin{array}{cc}
	\theta &  0\\
	0 & \eta
\end{array}\right), \;
\hat{J}_2 = \left(\begin{array}{cc}
0 & 1\\
-1& 0
\end{array}\right).
\end{eqnarray}
The effective Planck constant $\hbar_e$ is related with the position postion NC-parameter $\theta$ and momentum-momentum NC-parameter $\eta$ through the Planck constant $\hbar$ as
\begin{equation}
	\hbar_{e}=\hbar(1+\frac{\theta\eta}{4\hbar^2}).
\end{equation}
 In ~\eqref{NCcordefn} $X^T$ represents the matrix transposition of $X$.
The commutation relations for the usual commutative space 
\begin{equation}\label{Xdefn}
X=(\hat{X}_1, \hat{X}_2,\hat{X}_3,\hat{X}_4)^T=(\hat{x}_1, \hat{p}_1,\hat{x}_2,\hat{p}_2)^T
\end{equation}
are given by
\begin{equation}\label{Cspacecommutation}
	[	\hat{X}_{\alpha}, \hat{X}_{\beta}]=i\hbar \hat{J}_{\alpha\beta}= -\hbar(\hat{\Sigma}_y)_{\alpha\beta}, 
\end{equation}
with
\begin{equation}
	\hat{J}=\mbox{diag}(\hat{J}_2,\hat{J}_2),\;\; \hat{\Sigma}_y=\mbox{diag}(\hat{\sigma}_y, \hat{\sigma}_y).
\end{equation}
Unless otherwise specified,  in the present paper, we shall represent Pauli matrices as follows.
\begin{eqnarray}
	\hat{\sigma}_x =\left(\begin{array}{cc}
		0 & 1\\
		1& 0
	\end{array}\right),\;
	\hat{\sigma}_y =\left(\begin{array}{cc}
		0 & -i\\
		i& 0
	\end{array}\right),\;
	\hat{\sigma}_z =\left(\begin{array}{cc}
		1 & 0\\
		0& -1
	\end{array}\right).
\end{eqnarray}
The NC-space co-ordinates ($\tilde{X}$) are connected to the usual commutative space co-ordinates ($X$) through the Darboux transformation ($\hat{\Upsilon}_D$) given by the Bopp's shift
\begin{equation}\label{cncconnection}
	\tilde{X}=\hat{\Upsilon}_D X,
\end{equation}
with
\begin{eqnarray}
 \hat{\Upsilon}_D =\left(\begin{array}{cc}
	\hat{	\mathbb{I}}_2 & -\frac{1}{2\hbar}\hat{\Pi}_{\theta\eta}\hat{J}_2 \\
		\frac{1}{2\hbar}\hat{\Pi}_{\theta\eta}\hat{J}_2 & \hat{\mathbb{I}}_2 
	\end{array}\right),
\end{eqnarray}
where $\hat{\mathbb{I}}_n$ stands for $n\times n$ identity matrix.
From physical ground, since the position-position and momentum momentum non-commutativity will appear in much higher energy scale, we have $\theta<\hbar$ and $\eta< \hbar$, hence the determinant of $\hat{\Upsilon}_D $ is nonzero ($\Delta_{\Upsilon_D}\neq 0$); i.e., $\hat{\Upsilon}_D \in GL(4, R)$.
One can see that the symplectic matrix $\hat{J}$ is connected with the deformed symplectic matrix $\hat{\tilde{J}}$ through the following transformation via $\hat{\Upsilon}_D$.
\begin{equation}\label{JncJcconnection}
	\hbar_e \hat{\tilde{J}} = \hbar \hat{\Upsilon}_D \hat{J}\hat{\Upsilon}_D^T .
\end{equation}
Since the quantum mechanical formalism are well established in commutative space, it is customary to convert the NC-space system in usual commutative space system through ~\eqref{cncconnection} for computational purpose. 
For instance, one can define following matrix elements of a covariance matrix ($\hat{\tilde{\mathcal{V}}}_{nc}$) for NC-space co-ordinates.
\begin{equation}\label{vtildealphabetadefn}
	\tilde{\mathcal{V}}_{\alpha\beta}= \frac{1}{2}\langle \{ \hat{\tilde{X}}_\alpha , \hat{\tilde{X}}_\beta \}\rangle - \langle \hat{\tilde{X}}_\alpha \rangle \langle  \hat{\tilde{X}}_\beta \rangle ; \; \alpha,\beta \in \{1,2,3,4\},
\end{equation}
where the expectation value $\langle \hat{\chi}\rangle$ of the operator $\hat{\chi}$ is evaluated over the state in the usual commutative space variables. One can see that the  commutative space covariance matrix $\hat{\mathcal{V}}_c$, defined by
\begin{equation}\label{variancematrixdefninc}
	\mathcal{V}_{\alpha\beta}= \frac{1}{2}\langle \{ \hat{X}_\alpha , \hat{X}_\beta \}\rangle - \langle \hat{X}_\alpha \rangle \langle  \hat{X}_\beta \rangle ; \; \alpha,\beta \in \{1,2,3,4\},
\end{equation}
is connected with $\hat{\tilde{\mathcal{V}}}_{nc}$ through the following transformation via $\hat{\Upsilon}_D$.
\begin{equation}\label{VncVcconnection}
	\hat{\tilde{\mathcal{V}}}_{nc}= \hat{\Upsilon}_D  \hat{\mathcal{V}}_c\hat{\Upsilon}_D^T .
\end{equation}
Due to the fundamental commutation relations ~\eqref{Cspacecommutation} of commutative space, all the physically realizable variance matrix must satisfy the Robertson-Schr\"{o}dinger uncertainty principle (RSUP), which states that $\hat{\mathcal{V}}_c+\frac{i}{2} \hbar \hat{J}$ has to be positive definite; i.e.,  
\begin{equation}\label{RSUPc}
	\hat{\mathcal{V}}_c+\frac{i}{2}\hbar\hat{J} \ge 0 .
\end{equation}
Using~\eqref{JncJcconnection} and~\eqref{VncVcconnection}, one can see that  $\hat{\tilde{\mathcal{V}}}_{nc} + \frac{i}{2} \hbar_e \hat{\tilde{J}}$ is connected with  $\hat{\mathcal{V}}_c+\frac{i}{2} \hbar \hat{J}$ through $\hat{\Upsilon}_D$ as 
\begin{equation}
	\hat{\tilde{\mathcal{V}}}_{nc} + \frac{i}{2} \hbar_e \hat{\tilde{J}} = \hat{\Upsilon}_D \left( \hat{\mathcal{V}}_c+\frac{i}{2} \hbar \hat{J} \right) \hat{\Upsilon}_D^T .
	\end{equation}
A bonafide covariance matrix in NC-space has to satisfy the RSUP
\begin{equation}\label{RSUPnc}
\hat{\tilde{\mathcal{V}}}_{nc} + \frac{i}{2} \hbar_e \hat{\tilde{J}} \ge 0.
\end{equation}
	From ~\eqref{variancematrixdefninc}, one can note that $\hat{\mathcal{V}}_c$ can be written in the block form
	\begin{eqnarray}\label{cspacecovariancematrix}
		\hat{\mathcal{V}}_c=\left(\begin{array}{cc}
		V_{11} & V_{12}\\
			V_{12}^T & V_{22}
		\end{array}\right),
	\end{eqnarray}
A generic local transformation $\hat{S}_1 \bigoplus \hat{S}_2$, acts on $\hat{\mathcal{V}}_c$ as
\begin{equation}
	\hat{V}_{jj}\to \hat{S}_j \hat{V}_{jj}\hat{S}_j^T ,\; 	\hat{V}_{12}\to \hat{S}_1 \hat{V}_{12}\hat{S}_2^T;\; \mbox{with}\; \hat{S}_j \in Sp(2,\mathbb{R}),\;  j=1,2.
\end{equation}
One can identify that following four quantities are local  invariant with respect to transformation belonging to the $Sp(2,\mathbb{R})\bigotimes Sp(2,\mathbb{R}) \subset Sp(4,\mathbb{R})$.
\begin{eqnarray}
	\Delta_j=Det(V_{jj}),\; 	\Delta_{12}=Det(V_{12}), \; \Delta_{\mathcal{V}_c} =Det(\mathcal{V}_c)\\
	\tau_{v}=Trace(V_{11} J_2 V_{12} J_2 V_{22} J_2 V_{12}^T J_2).
\end{eqnarray}
 Williamson's theorem \cite{Williamson1,Williamson2} allows us to choose $\hat{S}_j$ such that 
\begin{eqnarray}
	\hat{\mathcal{V}}_c \to \left(\begin{array}{cc}
W_{11} & E_{12} \\
E_{12}^T & W_{22}
	\end{array}\right),
	\end{eqnarray}
where
\begin{equation}
	W_{jj}=\hat{S}_j \hat{V}_{jj}\hat{S}_j^T = d_j \hat{\mathbb{I}}_2,\; E_{12}= \hat{S}_1 \hat{V}_{12}\hat{S}_2^T.
\end{equation}
The eigenvalues ($d_j$) of $i\hat{J}_2\hat{V}_{jj} $ are the symplectic eigenvalues of $\hat{V}_{jj}$. In particular, for our present discussion
\begin{equation}
	d_j =\vert \sqrt{\Delta_j}\vert .
	\end{equation}
What follows from the above computation is that the variance matrix can be written into the normal form
\begin{eqnarray}
\hat{\mathcal{V}}_{nor} = \left(\begin{array}{cc}
 \sqrt{\Delta_1} \hat{\mathbb{I}}_2 & \mbox{diag}(\kappa_1,\kappa_2) \\
\mbox{diag}(\kappa_1,\kappa_2) &  \sqrt{\Delta_2} \hat{\mathbb{I}}_2
\end{array}\right),
\end{eqnarray}
where
\begin{equation}
\Delta_{12}=\kappa_1\kappa_2,\; \Delta_{\mathcal{V}_c}=( \sqrt{\Delta_1 \Delta_2}-\kappa_1^2) ( \sqrt{\Delta_1 \Delta_2}-\kappa_2^2). 
\end{equation}
Hence, one can conclude that RSUP ~\eqref{RSUPc} can be written as
\begin{equation}\label{covariancersupc}
	\Delta_1 \Delta_2 + (\frac{\hbar^2}{4}- \Delta_{12})^2 -\tau_{v}\ge \frac{\hbar^2}{4}( \Delta_1+\Delta_2).
\end{equation}
Under mirror reflection (Peres-Horodecki partial transpose) $\Delta_1$,  $\Delta_2$ and $\tau_v$ remains invariant; whereas, $\Delta_{12}$ flips sign. Therefore, the requirement that the covariance matrix of a separable state has to obey the following necessary condition.
\begin{equation}\label{separabilityc}
Ps=	\Delta_1 \Delta_2 + (\frac{\hbar^2}{4}-\vert \Delta_{12}\vert)^2 - \tau_{v}- \frac{\hbar^2}{4}( \Delta_1+\Delta_2) \ge 0,
\end{equation}
which turns out to be sufficient for all bipartite-Gaussian state.\\
 Since $\hat{\tilde{J}}$ is not symplectic and $\hat{\Upsilon}_D$ is not unitary, a straightforward generalization of ~\eqref{covariancersupc} for NC-space is not immediate, in spite of  the structural similarity of ~\eqref{Cspacecommutation} and ~\eqref{NCspacecommutation}. However, with the help of ~\eqref{cncconnection},  we can transform the NC-space system to commutative space, for which the general separability criterion ~\eqref{separabilityc} is at our aid. One can observe that Schur complement
 \begin{equation}
\hat{C}_{\Upsilon}=\hat{\mathbb{I}}_2 -\frac{1}{2\hbar}\hat{\Pi}_{\theta\eta}\hat{J}_2 \hat{\mathbb{I}}^{-1} (-\frac{1}{2\hbar})\hat{\Pi}_{\theta\eta}\hat{J}_2 =(1-\frac{\theta\eta}{4\hbar^2})\hat{\mathbb{I}}_2
 \end{equation} 
of $\mathbb{I}_2$ is nonsingular for all physically acceptable parameter values. In particular, 
\begin{equation}
\Delta_{C_{\Upsilon}}=	\mbox{Det}(\hat{C}_{\Upsilon}) = (1-\frac{\theta\eta}{4\hbar^2})^2\ge \frac{9}{16};\; \forall \theta\le\hbar,\eta\le\hbar.
\end{equation} 
In other words, $\hat{\Upsilon}_D^{-1}$ exists, and
\begin{eqnarray}
\hat{\Upsilon}_D^{-1} =\frac{1}{\sqrt{\Delta_{C_{\Upsilon}}}}  \left( \begin{array}{cc}
\hat{\mathbb{I}}_2 & \frac{1}{2\hbar}\hat{\Pi}_{\theta\eta} \hat{J}_2 \\
- \frac{1}{2\hbar}\hat{\Pi}_{\theta\eta} \hat{J}_2 & \hat{\mathbb{I}}_2
\end{array}
\right).
\end{eqnarray}
What follows is that for NC-space covariance matrix
\begin{eqnarray}
	\hat{\tilde{\mathcal{V}}}_{nc}=\left(\begin{array}{cc}
	\tilde{	V}_{11} & \tilde{ V}_{12}\\
		\tilde{V}_{12}^T & \tilde{V}_{22}
	\end{array}\right),
\end{eqnarray}
the generalized Peres-Horodecki criterion (Simon's criterion) may be studied with the help of the corresponding commuting space covariance matrix ~\eqref{cspacecovariancematrix}, with
\begin{eqnarray}
\hat{V}_{11} &=& \frac{1}{\Delta_{C_\Upsilon}} (\hat{\tilde{V}}_{11} + \frac{1}{2\hbar}(\hat{\tilde{V}}_{12}\hat{J}_2^T \hat{\Pi}_{\theta\eta}^T +  \hat{\Pi}_{\theta\eta} \hat{J}_2 \hat{\tilde{V}}_{12}^T ) + \frac{1}{4\hbar^2} \hat{\Pi}_{\theta\eta} \hat{J}_2 \hat{\tilde{V}}_{22} \hat{J}_2^T \hat{\Pi}_{\theta\eta}^T), \\
\hat{V}_{12} &=&   \frac{1}{\Delta_{C_\Upsilon}} (\hat{\tilde{V}}_{12} - \frac{1}{2\hbar}(\hat{\tilde{V}}_{11}\hat{J}_2^T \hat{\Pi}_{\theta\eta}^T -  \hat{\Pi}_{\theta\eta} \hat{J}_2 \hat{\tilde{V}}_{22} ) - \frac{1}{4\hbar^2} \hat{\Pi}_{\theta\eta} \hat{J}_2 \hat{\tilde{V}}_{12}^T \hat{J}_2^T \hat{\Pi}_{\theta\eta}^T), \\
\hat{V}_{22} &=&  \frac{1}{\Delta_{C_\Upsilon}} (\hat{\tilde{V}}_{22} - \frac{1}{2\hbar}(\hat{\tilde{V}}_{12}^T\hat{J}_2^T \hat{\Pi}_{\theta\eta}^T +  \hat{\Pi}_{\theta\eta} \hat{J}_2 \hat{\tilde{V}}_{12} ) + \frac{1}{4\hbar^2} \hat{\Pi}_{\theta\eta} \hat{J}_2 \hat{\tilde{V}}_{11} \hat{J}_2^T \hat{\Pi}_{\theta\eta}^T).
\end{eqnarray}
Explicitly written
\begin{eqnarray}
	\hat{V}_{11}= \frac{1}{\Delta_{C_\Upsilon}} \left(\begin{array}{cc}
	\tilde{\mathcal{V}}_{11}+\frac{\theta}{\hbar}\tilde{\mathcal{V}}_{14}+ \frac{\theta^2}{4\hbar^2}\tilde{\mathcal{V}}_{44} & 
	\tilde{\mathcal{V}}_{12}+ \frac{1}{2\hbar}(\theta \tilde{\mathcal{V}}_{24}-\eta \tilde{\mathcal{V}}_{13}) -\frac{\theta\eta}{4\hbar^2}\tilde{\mathcal{V}}_{34}\\
		\tilde{\mathcal{V}}_{12}+ \frac{1}{2\hbar}(\theta \tilde{\mathcal{V}}_{24}-\eta \tilde{\mathcal{V}}_{13}) -\frac{\theta\eta}{4\hbar^2}\tilde{\mathcal{V}}_{34} &
			\tilde{\mathcal{V}}_{22}-\frac{\eta}{\hbar}\tilde{\mathcal{V}}_{23}+ \frac{\eta^2}{4\hbar^2}\tilde{\mathcal{V}}_{33}
	\end{array}\right),\label{V11explicit} \\
	\hat{V}_{22}= \frac{1}{\Delta_{C_\Upsilon}} \left(\begin{array}{cc}
	\tilde{\mathcal{V}}_{33}-\frac{\theta}{\hbar}\tilde{\mathcal{V}}_{23}+ \frac{\theta^2}{4\hbar^2}\tilde{\mathcal{V}}_{22} & 
	\tilde{\mathcal{V}}_{34}- \frac{1}{2\hbar}(\theta \tilde{\mathcal{V}}_{24}-\eta \tilde{\mathcal{V}}_{13}) -\frac{\theta\eta}{4\hbar^2}\tilde{\mathcal{V}}_{12}\\
	\tilde{\mathcal{V}}_{34}- \frac{1}{2\hbar}(\theta \tilde{\mathcal{V}}_{24}-\eta \tilde{\mathcal{V}}_{13}) -\frac{\theta\eta}{4\hbar^2}\tilde{\mathcal{V}}_{12} &
	\tilde{\mathcal{V}}_{44}+\frac{\eta}{\hbar}\tilde{\mathcal{V}}_{14}+ \frac{\eta^2}{4\hbar^2}\tilde{\mathcal{V}}_{11}
\end{array}\right), \label{V22explicit}\\
	\hat{V}_{12}= \frac{1}{\Delta_{C_\Upsilon}} \left(\begin{array}{cc}
	\tilde{\mathcal{V}}_{13}-\frac{\theta}{2\hbar}(\tilde{\mathcal{V}}_{12}+\tilde{\mathcal{V}}_{34})- \frac{\theta^2}{4\hbar^2}\tilde{\mathcal{V}}_{24} & 
	\tilde{\mathcal{V}}_{14}+ \frac{1}{2\hbar}(\theta \tilde{\mathcal{V}}_{44}+\eta \tilde{\mathcal{V}}_{11}) +\frac{\theta\eta}{4\hbar^2}\tilde{\mathcal{V}}_{14}\\
	\tilde{\mathcal{V}}_{23}- \frac{1}{2\hbar}(\theta \tilde{\mathcal{V}}_{22}+\eta \tilde{\mathcal{V}}_{33}) +\frac{\theta\eta}{4\hbar^2}\tilde{\mathcal{V}}_{23} &
	\tilde{\mathcal{V}}_{24}-\frac{\eta}{2\hbar}(\tilde{\mathcal{V}}_{34}-\tilde{\mathcal{V}}_{12})- \frac{\eta^2}{4\hbar^2}\tilde{\mathcal{V}}_{13}
\end{array}\right), \label{V12explicit}
\end{eqnarray} 
where $\tilde{\mathcal{V}}_{\alpha\beta}$ are as defined in ~\eqref{vtildealphabetadefn}. The expressions ~\eqref{V11explicit}-~\eqref{V12explicit} can be used directly in ~\eqref{separabilityc} to determine the entanglement behavior, in particular the separability of the NC-coordinate degrees of freedom. In this paper, we have demonstrated the situation for a time-dependent anisotropic oscillator with a linear interaction term.  
\section{Time-dependent anisotropic oscillator with linear interaction with time-dependent parameters}
Along with bilinear Hamiltonians, linear field modes play important role in various domains.  For instance, in the Bosonization of one-dimensional cold atomic gases, collective atomic recoil laser using a Bose-Einstein condensate, linear field modes appear naturally \cite{linearfield1,linearfield2,linearfield3}. To study whether these linear terms  affect the covariance matrix, and hence the separability criterion, we consider 
the following Hamiltonian for an anisotropic oscillator with TD- parameters placed under an external TD- interaction, which is linear in coordinates, in two-dimensional NC-space.
\begin{equation}\label{nchamiltonian}
	\hat{H}_{nc}=\sum_{j=1}^{2} \left[\frac{1}{2m_j}\hat{\tilde{p}}_j^2 + \frac{1}{2}m_j \tilde{\omega}_j^2 \hat{\tilde{x}}_j^2 + \mathcal{E}_j \hat{\tilde{x}}_j\right],
\end{equation}
where the effective mass ($m(t)=(m_1(t),m_2(t))$)and frequency ($\omega(t)= (\omega_1(t), \omega_2(t))$) of the oscillator, as well as the parameters ( $\tilde{\mathcal{E}}(t) = ( \mathcal{E}_1(t),0, \mathcal{E}_2(t),0)$) of the external linear interaction are TD. Using the Bopp's shift ~\eqref{cncconnection}, we write the equivalent Hamiltonian $\hat{H}_c$ in commutative space as
\begin{equation}
	\hat{H}_c = \frac{1}{2}X^T\hat{\mathcal{H}}X + \mathcal{E}^T X,
\end{equation}
where
\begin{eqnarray}
	\hat{\mathcal{H}} =\left(\begin{array}{cc}
		\hat{C}& \hat{A}^T \\
		\hat{A} & \hat{B}
	\end{array}\right),\;\;
	\hat{A}=\left(\begin{array}{cc}
	0 & 2\nu_1\\
	-2\nu_2 & 0
\end{array}\right),\; 	\hat{B}=\left(\begin{array}{cc}
	\alpha_2	& 0 \\
	0	& \frac{1}{\mu_2}
\end{array}\right),\; 	\hat{C}=\left(\begin{array}{cc}
	\alpha_1 & 0\\
	0 & \frac{1}{\mu_1}
\end{array}\right),
\end{eqnarray}
and
\begin{eqnarray}
	\mathcal{E}=(\mathcal{E}_1,\frac{\theta}{2\hbar}\mathcal{E}_2, \mathcal{E}_2, - \frac{\theta}{2\hbar}\mathcal{E}_1)^T .
\end{eqnarray}
The TD - parameters $\alpha_j(t),\nu_j(t),\mu_j(t) $ are the concise expressions of the followings.
\begin{eqnarray}
	\frac{1}{\mu_j} &=& \frac{1}{m_j} + \frac{\theta^2}{4\hbar^2}m_l \tilde{\omega}_l^2 \vert \epsilon^{jl}\vert ,\\
	\alpha_j &=& m_j \tilde{\omega}_j^2 + \frac{\eta^2}{4\hbar^2 m_l} \vert \epsilon^{jl} \vert ,\\
	\nu_j &=& \frac{1}{4\hbar m_j}(\eta + m_1m_2 \theta \tilde{\omega}_l^2  \vert \epsilon^{jl} \vert ); \; j=1,2.
\end{eqnarray}
We have used the notation for  complete antisymmetric tensors (Levi-ci-Vita symbols) $\epsilon^{ij}$ with the convention $\epsilon^{12}=1$.\\
The intrinsic operator algebra ~\eqref{NCspacecommutation} provides an intrinsic symplectic structure 
\begin{equation}
	[\hat{H}_c,X]=\hbar \Sigma_y \hat{\mathcal{H}}X +\hbar \Sigma_y \mathcal{E} = -i\hbar \hat{\Omega} X +\hbar \Sigma_y \mathcal{E},
\end{equation}
which has to be preserved for any physically equivalent transformation. We wish to find normal co-ordinates that diagonalize our system. In other words,
we wish to find a similarity transformation that diagonalizes the nonsymmetric normal matrix $\hat{\Omega}$.   First we note that $\hat{\Omega}$ has four distinct imaginary eigenvalues
\begin{eqnarray}
	\lambda=\left\{ -i\lambda_1,i\lambda_1, -i\lambda_2,i\lambda_2 \right\}.
\end{eqnarray}
Where
\begin{equation}\label{lambda1lambda2vlues}
	\lambda_1 = \frac{1}{\sqrt{2}}\sqrt{b+\sqrt{\Delta}} ,\;
	\lambda_2 = \frac{1}{\sqrt{2}}\sqrt{b-\sqrt{\Delta}} ,
\end{equation}
with
\begin{eqnarray}
\Delta&=&b^2-4c,\;\;\;\;\;	b = \omega_1^2 + \omega_2^2 + 6\nu_1\nu_2 ,\\
	c &=& \omega_1^2  \omega_2^2 + 16 \nu_1^2 \nu_2^2 -4\nu_1^2 \omega_1^2 \mu_1/\mu_2 - 4\nu_2^2 \omega_2^2 \mu_2/\mu_1 .
\end{eqnarray}
If $u_j,\; j=1,2$ are left eigenvectors corresponding to the eigenvalue $-i\lambda_j$, then the left eigenvectors corresponding to $i\lambda_j$ are given by$	u_j^*$. Right eigenvectors are related to left eigenvectors by
$
	v_i= -\Sigma_y u_i^\dagger, \; (i=1,2).
$
Accordingly, the similarity transformation $\hat{Q}$ (as well as $\hat{Q}^{-1}$) can be constructed by arranging the  eigen-vectors column-wise. In particular,
\begin{equation}\label{Qdefn}
	\hat{Q}= (v_1,v_1^*,v_2,v_2^*), \;
	\hat{Q}^{-1} = (u_1^T,(u_1^*)^T,u_2^T,(u_2^*)^T).
\end{equation}
One can  verify that 
\begin{equation}\label{Qdagger}
	\hat{Q}^\dagger = -\hat{\Sigma}_z \hat{Q}^{-1} \hat{\Sigma}_y,
\;\mbox{with}\;
	\hat{\Sigma}_z = \mbox{diag}(\sigma_z,\sigma_z).
\end{equation}
For our computation purpose, we write the explicit form of the left eigenvectors as follows.
\begin{eqnarray}\label{leftevecOmega}
	u_i = \frac{1}{k_i} \left(-i\gamma_{i1} , \gamma_{i2}, \gamma_{i3}, i \gamma_{i4}
	\right), \; i=1,2.
\end{eqnarray}
with
\begin{eqnarray}
	\gamma_{i1} &=&\lambda_i \mu_1\mu_2(\lambda_i^2 -\omega_2^2 -4\nu_1\nu_2), \label{gamma1}\\
	\gamma_{i2}&=& \mu_2(\lambda_i^2 -\omega_2^2)+ 4\mu_1\nu_1^2, \\
	\gamma_{i3}&=& 2 \mu_1\mu_2 \nu_1(\lambda_i^2  -4\nu_1\nu_2)+ 2\nu_2\mu_2^2\omega_2^2,\\
	\gamma_{i4}&=&2\lambda_i (\mu_1\nu_1+\mu_2\nu_2). \label{gamma4}
\end{eqnarray}
$k_j$ are the normalization constants. We get   the following diagonal representation of $\hat{\Omega}$.
\begin{equation}
	\hat{\Omega}_D= \hat{Q}^{-1}\hat{\Omega}\hat{Q}=diag(-i\lambda_1,i\lambda_1,-i\lambda_2,i\lambda_2),
\end{equation}
One can define a normal co-ordinate 
\begin{equation}
	 \hat{A}=(\hat{A}_1,\hat{A}_2,\hat{A}_3,\hat{A}_4)^T = (\hat{a}_1,\hat{a}_1^\dagger,\hat{a}_2,\hat{a}_2^\dagger)^T,
\end{equation}
with the help of
	\begin{equation}
	X=QA,\; 	X^\dagger =  \hat{A}^\dagger (-\hat{\Sigma}_z \hat{Q}^{-1} \hat{\Sigma}_y) ,
\end{equation}
which satisfy the following algebra of annihilation ($\hat{a}_i$) and creation ($\hat{a}_i^{\dagger}$) operators.
\begin{equation}
	[\hat{A}_\alpha,\hat{A}_\beta]=i(\hat{\Sigma}_y)_{\alpha\beta} .
\end{equation}
 Accordingly, what follows is that the TD- Schr\"{o}dinger equation (TDSE)
$	\hat{H}_c \psi =i\hbar\frac{\partial\psi}{\partial t}$
is transformed in normal co-ordinates as 
\begin{equation}
	\frac{1}{2}A^\dagger \Sigma A \psi + \mathcal{E}^T QA\psi = i\hbar\frac{\partial\psi}{\partial t},
\end{equation}
where
\begin{equation}
\hat{	\Sigma} = i\hat{\Sigma}_z \hat{\Omega}_D =diag (\lambda_1,\lambda_1,\lambda_2,\lambda_2).
\end{equation}
That means we have an equivalent TDSE  
\begin{equation}\label{equivalentSE}
	\hat{H} \psi =i\hbar\frac{\partial\psi}{\partial t}
\end{equation}
with the following equivalent Hamiltonian for our system.
\begin{equation}\label{canonicalhamiltonian}
	\hat{H}(t)= \sum_{j=1}^{2}\left(\hat{a}_j^\dagger \hat{a}_j +\frac{1}{2}\right)\lambda_j + \mathcal{E}^T \hat{Q}\hat{A}.
\end{equation}

The Hamiltonian ~\eqref{canonicalhamiltonian} can be split into
\begin{equation}\label{Ht}
\hat{H}(t)=\hat{H}_0(t)+\hat{V}(t),
\end{equation}
where the eigenstates of 
\begin{equation}
	\hat{H}_0(t)=\sum_{i=1}^{2} \lambda_i (t)\hat{N}_i,\; \mbox{with}\; \hat{N}_i= \hat{a}_i^{\dagger}\hat{a}_i,
\end{equation}
are known. The additional TD- part is linear in field operators. In particular,
\begin{equation}
	\hat{V}(t)=g(t)+\sum_{i=1}^{2}\left( f_i(t) \hat{a}_i^{\dagger}+f_{i}^{*}(t) \hat{a}_i \right),
\end{equation}
with
\begin{eqnarray}
	f_j(t) &=&  \frac{1}{k_j}[\mathcal{E}_2 (\gamma_{j4}+\frac{\theta}{2\hbar}\gamma_{j1}) - i \mathcal{E}_1 (\gamma_{j2}+\frac{\theta}{2\hbar}\gamma_{j3})], \; j=1,2.\\
	g(t) &=& \frac{1}{2} (\lambda_1 +\lambda_2).
\end{eqnarray}
The real parameters $\gamma_{jk}$ in terms of the noncommutative parameters are given in Appendix (Equation ~\eqref{gamma1}-~\eqref{gamma4}).

Our first task is to find out the solutions for ~\eqref{Ht}. Since $\lambda_j(t)$ and $f_j(t)$ are TD, we shall adopt Lewis-Riesenfeld phase-space invariant method to solve the system. 
\section{Lewis-Riesenfeld  invariant operator and the solution of Schr\"{o}dinger equation }
We shall apply the Lewis-Riesenfeld method as prescribed in \cite{Lewis4}. Let us assume that the eigenvalue equation of the TD-operator
\begin{equation}
	\hat{\mathcal{O}}(t)={\displaystyle \prod_{j=1}^{2} \hat{\mathcal{O}}_j(t)},\;\mbox{with}\; \hat{\mathcal{O}}_j(t) = \exp\left(\mu_j (t)\hat{N}_j\right),
\end{equation}
are known for arbitrary complex TD-parameters $\mu_j(t)\; (j=1,2)$.
 However $\hat{\mathcal{O}}(t)$ is  not invariant associated with the total Hamiltonian $\hat{H}(t)$. In particular,
\begin{equation}\label{Thetat}
\hat{\Theta}(t)=\partial_t \hat{\mathcal{O}}+(i\hbar)^{-1}[\hat{\mathcal{O}},\hat{H}]\neq 0.
\end{equation}
We readily see that if $\vert \psi(t)\rangle$ is a solution of the TDSE ~\eqref{equivalentSE} associated with $\hat{H}(t)$, then $\hat{\mathcal{O}}(t) \vert \psi(t)\rangle$ is a solution of TDSE for the TD-Hamiltonian 
\begin{equation}
\hat{	\tilde{H}}(t)=\hat{H}(t)+ i\hbar \Theta(t)\hat{\mathcal{O}}^{-1}(t).
\end{equation}
We introduce a TD-unitary transformation $\hat{\Lambda}(t)$, which relates the solutions of both TDSE, associated with $\hat{H}(t)$ and $\hat{\tilde{H}}(t)$, leading to
\begin{equation}
	i\hbar \partial_t[\hat{\Lambda}(t)\hat{\mathcal{O}}(t)\vert \psi(t)\rangle] =\hat{\tilde{\mathcal{H}}}(t)[\hat{\Lambda}(t)\hat{\mathcal{O}}(t)\vert \psi(t)\rangle],
\end{equation}
where
\begin{equation}
\hat{\tilde{\mathcal{H}}}(t)= \hat{\Lambda}(t) \hat{H}(t) \hat{\Lambda}^{-1}(t) + i\hbar \hat{\Lambda}(t) \hat{\Theta}(t)\hat{\mathcal{O}}^{-1}(t) \hat{\Lambda}^{-1}(t) + i\hbar \partial_t[\hat{\Lambda}(t)]\hat{\Lambda}^{-1}(t).
\end{equation}
We demand $\hat{\tilde{\mathcal{H}}}(t)=\hat{H}(t)$, which implies
\begin{equation}\label{LambdaHcommut}
	\partial_t\hat{\Lambda}(t) -\frac{1}{i\hbar}[\hat{H}(t),\hat{\Lambda}(t)]=-\hat{\Lambda}(t)\hat{\Theta}(t) \hat{\mathcal{O}}^{-1}(t).
\end{equation}
One can verify that $\hat{\Lambda}(t)\hat{\mathcal{O}}(t)\vert \psi(t)\rangle$ is a solution of TDSE associated with $\hat{H}(t)$. Using ~\eqref{Thetat} and ~\eqref{LambdaHcommut}, we find that $\hat{\mathcal{I}}(t)=\hat{\Lambda}(t)\hat{\mathcal{O}}(t)$ is a TD-invariant for $\hat{H}(t)$. In other words,
\begin{equation}
	\partial_t [\hat{\Lambda}(t)\hat{\mathcal{O}}(t)] + ( i\hbar)^{-1} [ \hat{\Lambda}(t)\hat{\mathcal{O}}(t), \hat{H}(t)]=0.
\end{equation}
According to Lewis-Riesenfeld (LR) theorem, if  $\vert n,t\rangle$ is an eigenstate of $\hat{\mathcal{I}}(t)$, then $\vert n,t\rangle e^{i\phi(t)}$ is a solution of TDSE for $H(t)$, for some TD-phase factor $\phi(t)$, which can be computed from consistency condition. \\
Accordingly, our first aim to compute $\Theta(t)$. Then we have to take a consistent ansatz for $\Lambda$ so that a LR-invariant operator $\hat{\mathcal{I}}(t)$ can be constructed. The construction of eigenstates of $\hat{\mathcal{I}}(t)$, along with phase-factor $\phi(t)$ will lead to our desired solution.\\
Since $[\hat{H}_0,\hat{\mathcal{O}}_i]=0$, equation~\eqref{Thetat} is reduced to
\begin{equation}\label{Thetatv}
	\hat{\Theta}(t)=\partial_t \hat{\mathcal{O}}+(i\hbar)^{-1}[\hat{\mathcal{O}},\hat{V}].
\end{equation}
Using Kubo's identity
\begin{equation}
	[\hat{A},e^{\mu \hat{B}}]=-\int_{0}^{-\mu} e^{(\mu+u)\hat{B}}[\hat{A},\hat{B}]e^{-u\hat{B}}du,
\end{equation}
we have the following identities.
\begin{eqnarray}
	\begin{array}{cc}
			\left[\hat{a}_i,\hat{\mathcal{O}}_j\right]=(e^{\mu_j }-1) \hat{\mathcal{O}}_j\hat{a}_j\delta_{ij}, & \hat{\mathcal{O}}_j \hat{a}_j \hat{\mathcal{O}}_j^{-1}=e^{-\mu_j}\hat{a}_j, \\
			\left[\hat{a}_i^\dagger,\hat{\mathcal{O}}_j\right]=(e^{-\mu_j }-1)\hat{\mathcal{O}}_j\hat{a}_j^\dagger \delta_{ij} , & 
			\hat{\mathcal{O}}_j \hat{a}_j^\dagger \hat{\mathcal{O}}_j^{-1}=e^{\mu_j}\hat{a}_j^\dagger,
		\end{array}
\end{eqnarray}
which leads to
\begin{eqnarray}
	\hat{\Theta}(t) \hat{\mathcal{O}}^{-1}= \sum_{j=1}^{2} \dot{\mu}_j (t)\hat{a}_j^{\dagger}\hat{a}_j +(i\hbar)^{-1}  \sum_{j=1}^{2} [f_j(e^{\mu_j}-1)\hat{a}_j^\dagger + f_j^*(e^{-\mu_j}-1)\hat{a}_j ].
\end{eqnarray}
We now consider following ansatz for $\hat{\Lambda}(t)$ through the displacement operators.
\begin{equation}
	\hat{\Lambda}(t)={\displaystyle \prod_{j=1}^{2}\hat{\Lambda}_j(t)},\; \mbox{with}\; \hat{\Lambda}_j(t)= \hat{\mathcal{D}}_j(\beta_j) =e^{\beta_j\hat{a}_j^\dagger -\beta_j^* \hat{a}_j}.
\end{equation}
The commutation relations and transformation rules for the annihilation, creation and number operators with $	\hat{\Lambda}(t)$ are the followings.
\begin{eqnarray}
	\left[\hat{a}_k,\hat{\Lambda}(t)\right] &=&\beta_k\hat{\Lambda}(t), \label{alambdacommut}
	\left[\hat{a}_k^\dagger,\hat{\Lambda}(t)\right]=\beta_k^* \hat{\Lambda}(t), \label{adaggerlambdacommut} \\
	\left[\hat{a}_k^\dagger\hat{a}_k,\hat{\Lambda}(t)\right]&=&(\beta_k\hat{a}_k^\dagger + \beta_k^*\hat{a}_k -\vert \beta_k\vert^2)\hat{\Lambda}(t). \label{numberlambdacommut}\\
	\hat{ \Lambda}^\dagger \hat{a}_j \hat{ \Lambda} &=& \hat{a}_j  +\beta_j ,\;
	\hat{ \Lambda}^\dagger \hat{a}_j^\dagger \hat{ \Lambda} = \hat{a}_j^\dagger  +\beta_j^* , j=1,2. \label{atransformlambda}
\end{eqnarray}
Using ~\eqref{alambdacommut},~\eqref{adaggerlambdacommut} and ~\eqref{numberlambdacommut} we get
\begin{equation}
	[\hat{H}(t),\hat{\Lambda}(t)] = \sum_{k=1}^{2} [\omega_k(\beta_k \hat{a}_k^\dagger + \beta_k^* \hat{a}_k) + (f_k\beta_k^* + f_k^*\beta_k)-\omega_k \vert \beta_k\vert^2 ] \hat{\Lambda}.
\end{equation}
Moreover
\begin{equation}\label{Lambdaderivative}
	\partial_t\hat{\Lambda}= \sum_{k=1}^{2} [\dot{\beta}_k\hat{a}_k^\dagger -\dot{\beta}_k^*\hat{a}_k +\beta_k\dot{\beta}_k^* -\frac{1}{2}\partial_t \vert \beta_k\vert^2]\hat{\Lambda}.
\end{equation}
And
\begin{eqnarray}
	\hat{\Lambda}\hat{\Theta}\hat{\mathcal{O}}^{-1} &=&\sum_{k=1}^{2} [\dot{\mu}_k (\hat{a}_k^\dagger \hat{a}_k -\beta_k\hat{a}_k^\dagger -\beta_k^*\hat{a}_k +\vert \beta_k\vert^2) + (i\hbar)^{-1}\{f_k(e^{\mu_k}-1)(\hat{a}_k^\dagger -\beta_k^*)  \nonumber \\
	&& + f_k^*(e^{-\mu_k}-1)(\hat{a}_k -\beta_k)\}]\hat{\Lambda}.
\end{eqnarray}
Equation ~\eqref{LambdaHcommut} provides  consistency conditions through the following first order differential equations of the displacement parameters $\beta_k(t)$ and the unknowns $\mu_k(t)$.
\begin{eqnarray}
	i\hbar\dot{\beta}_j &=&\lambda_j\beta_j+i\hbar \dot{\mu}_j\beta_j- f_j(e^{\mu_j}-1),\; j=1,2. \label{beta1dot}\\
		-i\hbar\dot{\beta}_j^* &=& \lambda_j\beta_j^*+ i\hbar \dot{\mu}_j\beta_j^*-f_j^*(e^{-\mu_j}-1),\; j=1,2. \label{beta1stardot}\\
	\dot{\mu}_1 &=& 	\dot{\mu}_2=0 , \label{mu1dot}
\end{eqnarray}
along with  the constraint
\begin{eqnarray}\label{constraintparameter}
	i\hbar\sum_{k=1}^{2}(\beta_k\dot{\beta}_k^*-\frac{1}{2}\partial_t \vert \beta_k \vert^2) &=& \sum_{k=1}^{2}(f_k\beta_k^* +f_k^*\beta_k -\lambda_k\vert \beta_k\vert^2 +(-1)^ki\hbar \dot{\mu}_k\vert \beta_k\vert^2 \nonumber \\
	&& + f_k(e^{\mu_k}-1)\beta_k^* + f_k^*(e^{-\mu_k}-1)\beta_k).
	\end{eqnarray}
From ~\eqref{mu1dot}  we have $\mu_1$ and $\mu_2$ are constants. On the other hand, the comparison of ~\eqref{beta1dot}  with complex conjugation of ~\eqref{beta1stardot} leads to
\begin{equation}\label{mu1mu2purelyimaginary}
	\mu_1=i\mu_{10},\; 	\mu_2=i\mu_{20}, \;\mbox{with}\; \mu_{10}, \mu_{20}\in \mathbb{R}.
\end{equation}
The set of independent equations are then
\begin{equation}
	i\hbar\dot{\beta}_j =\lambda_j\beta_j -f_j(e^{i\mu_{j0}}-1),\;j=1,2.\label{beta1dotreduced}
\end{equation}
Using ~\eqref{mu1mu2purelyimaginary} in the constraint equation ~\eqref{constraintparameter} we have
\begin{eqnarray}
	\sum_{j=1}^{2}[\lambda_j\vert \beta_j\vert^2 +2\Re(f_j\beta_j^*)+\hbar\Im(\dot{\beta}_j \beta_j^*)] =0.\\
		\mu_{j0} = (2n_j+1)\pi ;\;n_j=0,\pm 1,\pm 2, \pm 3........\label{muj0int}
\end{eqnarray}
The notation $\Re(z)$ and $\Im(z)$ stands for the real and imaginary part of $z$, respectively. 
Given time-dependent parameters ($\lambda_j(t),f_j(t),\; j=1,2$) one can solve ~\eqref{beta1dotreduced}. In particular,
\begin{equation}
	\beta_j(t)=\beta_{j0} e^{-i\Omega_j/\hbar} - \frac{2i}{\hbar} e^{-i\Omega_j/\hbar} \int^{t} f_j(\tau) e^{i\Omega_j(\tau)/\hbar}d\tau , \; j=1,2.
\end{equation}
Where $\beta_{j0}$ are integration constants, and
\begin{equation}
	 \Omega_j(t)=\int^{t}\lambda_j(\tau)d\tau.
\end{equation}
For each choice of a pair of integers $(l_1,l_2)$ in ~\eqref{muj0int}, we have a Lewis Riesenfeld invariant operator, which reads
\begin{equation}
\hat{I}_{l_1,l_2}(t)= \hat{\Lambda}(t)\hat{\mathcal{O}}=\hat{\mathcal{D}}(\beta_1) \hat{\mathcal{D}}(\beta_2)e^{(2l_1+1)i\pi \hat{N}_1 + (2l_2+1)i\pi \hat{N}_2}.
\end{equation}
One can readily identify that the eigenstates $\{ \vert n_1,n_2;\beta_1,\beta_2 \rangle \}$ of the invariant operator $\hat{I}_{l_1,l_2}$ are the displaced number states
\begin{equation}
 \vert n_1,n_2;\beta_1,\beta_2 \rangle = e^{(2l_1+1)i\pi n_1 + (2l_2+1)i\pi n_2}\hat{\mathcal{D}}(\beta_1) \hat{\mathcal{D}}(\beta_2) \vert n_1,n_2\rangle .
\end{equation}
\subsection{Phase factor}
Time-dependent phase-factors $\phi_{n_1,n_2}(t)$, associated with the solutions ($	\vert \psi(t)\rangle = \sum_{n_1,n_2} c_{n_1,n_2}  \vert n_1,n_2;\beta_1,\beta_2 \rangle e^{i\phi_{n_1,n_2}(t)}$) of TDSE are given by
\begin{eqnarray}
	\phi_{n_1,n_2}(t) =\frac{1}{\hbar} \int_{0}^{t} \langle n_1,n_2\vert \hat{ \Lambda}^\dagger  (i\hbar \frac{\partial}{\partial t} -\hat{H}(\tau)) \hat{\Lambda } \vert n_1,n_2\rangle d\tau .
\end{eqnarray}
We can split phase-factors into a geometric part ($\phi_{n_1,n_2}^{(g)}$) and a dynamic part ($\phi_{n_1,n_2}^{(d)}$) as 
\begin{eqnarray}
\phi_{n_1,n_2}^{(g)}(t) =\frac{1}{\hbar} \int_{0}^{t} \langle n_1,n_2\vert \hat{ \Lambda}^\dagger  i\hbar \frac{\partial}{\partial t} \hat{\Lambda } \vert n_1,n_2\rangle d\tau ,\label{geometricphasedefn}\\
\phi_{n_1,n_2}^{(d)}(t) =-\frac{1}{\hbar} \int_{0}^{t} \langle n_1,n_2\vert \hat{ \Lambda}^\dagger \hat{H} \hat{\Lambda } \vert n_1,n_2\rangle d\tau . \label{dynamicalphasedefn}
\end{eqnarray}
Using ~\eqref{Lambdaderivative} along with the unitarity of $\hat{\Lambda}$ we can simplify ~\eqref{geometricphasedefn}. Moreover, the  identities ~\eqref{atransformlambda}
 reduce $\phi_{n_1,n_2}^{(g)}(t)$ to
\begin{equation}\label{geometricphaseform}
\phi_{n_1,n_2}^{(g)}(t)=	\frac{i}{2}\sum_{k=1}^{2}\int_{0}^{t} \left(\frac{\partial \beta_k}{\partial \tau}\beta_k^* -\beta_k \frac{\partial \beta_k^*}{\partial \tau}\right)d\tau =-\sum_{k=1}^{2}\int_{0}^{t} \Im \left(\frac{\partial \beta_k}{\partial\tau}\beta_k^*\right)d\tau,
\end{equation}
where we have used the orthonormality conditions $(\langle n_i,n_k\vert n_j,n_l\rangle =\delta_{ij}\delta_{kl})$ of number states. \\
For dynamical phase, we first note the following transformation rule of $\hat{H}$ under $\hat{\Lambda}$.
\begin{equation}\label{HtransformLambda}
\hat{ \Lambda}^\dagger \hat{H} \hat{ \Lambda} = \hat{H} + \sum_{j=1}^{2} \lambda_j(\beta_j \hat{a}_j^\dagger + \beta_j^*\hat{a}_j) + (\lambda_j\vert \beta_j \vert^2 +f_j \beta_j^* + f_j^*\beta_j).
\end{equation}
~\eqref{HtransformLambda} simplifies ~\eqref{dynamicalphasedefn}, which enables us to write 
\begin{equation}\label{dynamicphaseform}
\phi_{n_1,n_2}^{(d)}(t)= -\frac{1}{\hbar} \int_{0}^{t} \left[g(\tau) + \sum_{k=1}^{2} (\lambda_k n_k + \lambda_k\vert \beta_k\vert^2 + f_k\beta_k^* + f_k^*\beta_k) \right]d\tau .
\end{equation}
From ~\eqref{geometricphaseform} and ~\eqref{dynamicphaseform}, we see that time dependent phase factors
\begin{equation}
	\phi_{n_1,n_2}(t)= \phi_{n_1,n_2}^{(g)}(t) + \phi_{n_1,n_2}^{(d)}(t)
\end{equation}
are real functions of time. Thus we have a set of solutions for the concerned TDSE. Now with these solutions, one can construct the covariance matrix and study the separability criterion, which is discussed in next section.
\section{ Separability criterion for the system}
Since the connection between the NC-space co-ordinates and usual commutative space co-ordinate are given by
\begin{equation}
	\hat{\tilde{X}}_k= \sum_{j=1}^{4}\upsilon_{kj}\hat{X}_j,
\end{equation}
one can write
\begin{eqnarray}
	\langle \hat{\tilde{X}}_k \rangle =\sum_{j=1}^{4}\upsilon_{kj} \left(\sum_{l=1}^{4}Q_{jl}\langle \hat{A}_l \rangle \right),
\end{eqnarray}
where $\upsilon_{kj}$ is the $kj^{th}$-element of $\hat{\Upsilon_D}$ as specified in ~\eqref{cncconnection}, and $Q_{\alpha\beta}$ is the $\alpha\beta^{th}$ element of the similarity transformation matrix $\hat{Q}$ as specified in ~\eqref{Qdefn}. Thus the
expectation value of $\hat{A}_l$ over the state $\vert n_1,n_2;\beta_1,\beta_2 \rangle e^{i\phi_{n_1,n_2}} = \hat{\Lambda}\vert n_1,n_2\rangle e^{i\phi_{n_1,n_2}}$ can be written as
\begin{eqnarray}
	\langle \hat{A}_l \rangle &=&\langle n_1,n_2\vert \hat{\Lambda}^\dagger \hat{A}_l \hat{\Lambda}\vert n_1,n_2\rangle ;\;\; \because \phi_{n_1,n_2}(t)\in \mathbb{R}. \nonumber \\
	&=& \langle n_1,n_2\vert (\hat{A}_l +\tilde{\beta}_l) \vert n_1,n_2\rangle ,
\end{eqnarray}
where,
\begin{equation}
	\tilde{\beta}=(	\tilde{\beta}_1,	\tilde{\beta}_2, 	\tilde{\beta}_3, 	\tilde{\beta}_4)=(\beta_1,\beta_1^*, \beta_2,\beta_2^*).
\end{equation}
Since $\hat{A}_l$ are just annihilation or creation operators, and  the number states are orthonormalized, we have $\langle n_1,n_2\vert \hat{A}_l \vert n_1,n_2\rangle =0$. 
Therefore
\begin{equation}\label{xalphatildeexpectation}
	\langle \hat{\tilde{X}}_\alpha \rangle = \sum_{j=1}^{4}\upsilon_{\alpha j} \left(\sum_{l=1}^{4}Q_{jl} \tilde{\beta}_l\right).
\end{equation}
Similarly,
\begin{eqnarray}\label{xatildexbtildeexpectation}
	\langle \{\hat{\tilde{X}}_\alpha, \hat{\tilde{X}}_\beta \}\rangle =2 \sum_{j=1}^{4}\sum_{l=1}^{4} \upsilon_{\alpha j} \upsilon_{\beta l} \Re (Q_{jl}^{(13)}) + 2 \langle \hat{\tilde{X}}_\alpha \rangle \langle \hat{\tilde{X}}_\beta \rangle ,
\end{eqnarray}
where
\begin{equation}
	Q_{jl}^{(13)} = \bar{Q}_{lj}^{(13)} = n_1 \bar{Q}_{j1} Q_{l1} + (1+n_1)  Q_{j1} \bar{Q}_{l1} +  n_2 \bar{Q}_{j3} Q_{l3} + (1+n_2)  Q_{j3} \bar{Q}_{l3} .
\end{equation}
Here the notation $\bar{z}$ denotes the complex conjugate of the complex number $z$.\\
Using ~\eqref{xalphatildeexpectation} and ~\eqref{xatildexbtildeexpectation} in ~\eqref{vtildealphabetadefn}, we see that the covariance matrix elements are reduced to
\begin{equation}\label{varianceelements}
\tilde{	\mathcal{V}}_{\alpha\beta}=\sum_{j=1}^{4}\sum_{l=1}^{4} \upsilon_{\alpha j} \upsilon_{\beta l} \Re (Q_{jl}^{(13)}) ,
\end{equation}
which are independent of the displacement parameters $\beta_j$. However, $\tilde{	\mathcal{V}}_{\alpha\beta}$ depends on the matrix elements of $\hat{Q}$. In other words, the covariance matrix depends on the noncommutative parameters, as well as the time-dependent parameters $m_j, \tilde{\omega}_j$ corresponding to the system. On the other hand, we see that the contributions of the external influence $\tilde{\mathcal{E}}$ are manifested only through $\tilde{\beta}$, which has no effect in the covariance matrix. Therefore the entanglement property (if any) between the NC space (NCS) coordinate degrees of freedom (DOF) arises only due to the NC-space parameters $\upsilon_{ij}$. Moreover, an arbitrary linear displacement does not affect the separability criterion for the system. \\
The separability criterion will provide a restriction on the parameter values, for which the NCS DOF will be separable. For completeness, what follows is an inequality that has to be satisfied by the parameters of the separability of NCS DOF.
\\
Using the explicit forms of the elements of $\hat{Q}$, we get the following forms of $Q_{jl}^{(13)}$.
\begin{eqnarray}
Q_{11}^{(13)} &=& (2n_1+1)\gamma_{12}^2 + (2n_2+1 )\gamma_{22}^2 =q_{11}, \label{q11}\\
Q_{12}^{(13)} &=& i(\gamma_{11}\gamma_{12}+ \gamma_{21}\gamma_{22})=\bar{Q}_{21}^{(13)} =iq_{12},\\
Q_{13}^{(13)} &=&  i(\gamma_{12}\gamma_{14}+ \gamma_{22}\gamma_{24})=\bar{Q}_{31}^{(13)}=iq_{13},\\
Q_{14}^{(13)} &=& -(2n_1+1)\gamma_{12}\gamma_{13} - (2n_2+1 )\gamma_{22}\gamma_{23} = \bar{Q}_{41}^{(13)} =q_{14},\\
Q_{22}^{(13)} &=& (2n_1+1)\gamma_{11}^2 + (2n_2+1 )\gamma_{21}^2 =q_{22} ,\\
Q_{23}^{(13)} &=& (2n_1+1)\gamma_{11}\gamma_{14} + (2n_2+1 )\gamma_{21}\gamma_{24}  = \bar{Q}_{32}^{(13)}=q_{23},\\
Q_{24}^{(13)} &=&  i(\gamma_{11}\gamma_{13}+ \gamma_{21}\gamma_{23}) = \bar{Q}_{42}^{(13)}=iq_{24},\\
Q_{33}^{(13)} &=&  (2n_1+1)\gamma_{14}^2 + (2n_2+1 )\gamma_{24}^2 =q_{33} ,\\
Q_{34}^{(13)} &=&  i(\gamma_{13}\gamma_{14}+ \gamma_{23}\gamma_{24}) = \bar{Q}_{43}^{(13)}=iq_{34},\\
Q_{44}^{(13)} &=&  (2n_1+1)\gamma_{13}^2 + (2n_2+1 )\gamma_{23}^2 =q_{44}. \label{q44}
\end{eqnarray}
Here we have defined $q_{\alpha\beta}\in \mathbb{R}$ for convenience. The covariance matrix elements in NC-space now reads
\begin{eqnarray}
			\tilde{\mathcal{V}}_{11} &=& q_{11} -\frac{\theta}{\hbar} q_{14} + \frac{\theta^2}{4\hbar^2} q_{44}, \;	\tilde{\mathcal{V}}_{12} = 0, \;	\tilde{\mathcal{V}}_{13} = 0, \;	\tilde{\mathcal{V}}_{14} = \frac{\hbar_e}{\hbar} q_{14} - \frac{\eta}{2\hbar} q_{11} - \frac{\theta}{2\hbar} q_{44}, \nonumber \\
			\tilde{\mathcal{V}}_{21} &=& 0, \;	\tilde{\mathcal{V}}_{22} = q_{22} + \frac{\eta}{\hbar} q_{23} + \frac{\eta^2}{4\hbar^2} q_{33}, \;	\tilde{\mathcal{V}}_{23} = \frac{\hbar_e}{\hbar} q_{23} + \frac{\theta}{2\hbar} q_{22} + \frac{\eta}{2\hbar} q_{33}, \;	\tilde{\mathcal{V}}_{24} = 0, \nonumber \\
				\tilde{\mathcal{V}}_{31} &=& 0, \; 	\tilde{\mathcal{V}}_{32}=	\tilde{\mathcal{V}}_{23}  , \; 	\tilde{\mathcal{V}}_{33} = q_{33} + \frac{\theta}{\hbar} q_{23} + \frac{\theta^2}{4\hbar^2} q_{22}, \; 	\tilde{\mathcal{V}}_{34}=0, \nonumber \\
				  	\tilde{\mathcal{V}}_{41} &=&	\tilde{\mathcal{V}}_{14}, \; 	\tilde{\mathcal{V}}_{42}=0, \; 	\tilde{\mathcal{V}}_{43}=0,\; 
				  		\tilde{\mathcal{V}}_{44} = q_{44} -\frac{\eta}{\hbar} q_{14} + \frac{\eta^2}{4\hbar^2} q_{11}. \label{explicitformvtildealphabeta}
\end{eqnarray}
In particular, the covariance matrix reads
\begin{eqnarray}
	\tilde{\mathcal{V}}=  \left(\begin{array}{cccc}
q_{11} & 0 & 0 & q_{14}\\
0 & q_{22} & q_{23} & 0 \\
0 & q_{23} & q_{33} & 0\\
q_{14} & 0 & 0 & q_{44}
	\end{array}\right) + 
\frac{\theta}{\hbar} \left(\begin{array}{cccc}
-q_{14} & 0 & 0 & -\frac{q_{44}}{2}\\
0 & 0 & \frac{q_{22}}{2} & 0 \\
0 & \frac{q_{22}}{2} & q_{23} & 0\\
-\frac{q_{44}}{2} & 0 & 0 & 0
\end{array}\right)+ 
\frac{\eta}{\hbar} \left(\begin{array}{cccc}
0 & 0 & 0 & -\frac{q_{11}}{2} \\
0 & q_{23} & \frac{q_{33}}{2} & 0 \\
0 & \frac{q_{33}}{2} & 0 & 0\\
- \frac{q_{11}}{2} & 0 & 0 & -q_{14}
\end{array}\right) \nonumber \\ 
+
\frac{\theta^2}{4\hbar^2}\left(\begin{array}{cccc}
	q_{44} & 0 & 0 & 0 \\
	0&0&0&0\\
	0& 0 & q_{22}& 0\\
	0 & 0&0&0
\end{array}\right)
+
\frac{\eta^2}{4\hbar^2}\left(\begin{array}{cccc}
	0 & 0 & 0 & 0 \\
	0& q_{23}&0&0\\
	0& 0 & 0& 0\\
	0 & 0&0&q_{11}
\end{array}\right)
+ \frac{\theta\eta}{4\hbar^2} \left(\begin{array}{cccc}
 0 & 0 & 0 & q_{14} \\
 0 & 0 & q_{23} & 0 \\
 0 & q_{23} & 0 & 0 \\
 q_{14} & 0 & 0 & 0
\end{array} \right). 
\end{eqnarray}

Using the expressions ~\eqref{explicitformvtildealphabeta} in  ~\eqref{V11explicit},~\eqref{V22explicit} and ~\eqref{V12explicit}, surprisingly we obtain the following simple form for $\hat{V}_{ij}$.  
\begin{eqnarray}\label{varianceblockincommcoordinate}
	\hat{V}_{11}= \left(\begin{array}{cc}
q_{11} & 0 \\
0 & q_{22}
	\end{array}\right),\; 
	\hat{V}_{22}= \left(\begin{array}{cc}
	q_{33} & 0 \\
	0 & q_{44}
\end{array}\right),\;
	\hat{V}_{12}= \left(\begin{array}{cc}
	0 & q_{14} \\
	q_{23} & 0
\end{array}\right).
\end{eqnarray}
One can now use ~\eqref{varianceblockincommcoordinate} in ~\eqref{separabilityc} to study the separability of the NC-space coordinate degrees of freedom.  Since the generalized Peres-Horodecki criterion (Simon's condition) is a sufficient condition for Gaussian states, from now on we shall be using $n_1=n_2=0$. For convenience, we shall stick to the natural unit system, for which $\hbar=1$.\\
Interestingly, one can see that $Ps$ has the commutative space ($\theta\to 0,\eta\to 0$) limit
\begin{equation}
\lim\limits_{\theta\to 0,\eta\to 0,\hbar\to 1}Ps=\frac{1}{16}- \frac{1}{4}m_1^2 m_2^4 \omega_1^2 (\omega_2^2-\omega_1^2)^4,
\end{equation}
which is always positive (thus separable) for an isotropic oscillator. However, for an anisotropic oscillator, it might be negative for specific values of the parameters. For instance, for $m_2=\omega_2=1$ and $m_1\omega_1=1$, separable states are obtained for
\begin{equation}
	1-\frac{1}{\sqrt{2}}\le \omega_1^2 \le 1+\frac{1}{\sqrt{2}}.
\end{equation}
\section{Toy Model: Isotropic oscillator with only spatial noncommutativity }
For illustration, here we consider only spatial noncommutativity
\begin{equation}\label{toyncalgebra}
	[\hat{\tilde{x}}_1,\hat{\tilde{x}}_2]=i\theta(t),\; [\hat{\tilde{p}}_1,\hat{\tilde{p}}_2]=0.
\end{equation} 
Let us consider an isotropic oscillator placed on that background noncommutative space. For simplicity, without loss of generality, we consider the following parameter values for the isotropic oscillator.
\begin{equation}
	m_1=m_2=1,\; \tilde{\omega}_1=\tilde{\omega}_2=1, \; \hbar=1. 
\end{equation}
Now using explicit forms ~\eqref{lambda1lambda2vlues}, ~\eqref{gamma1}-~\eqref{gamma4}, and ~\eqref{q11}-~\eqref{q44} in the covariance matrix blocks ~\eqref{varianceblockincommcoordinate}, we can compute $\Delta_{1},\Delta_2,\Delta_{12},\tau_v$ in straightforward manner. Since the sign of $\Delta_{12}$ is crucial, we write it explicitly for the isotropic case, as follows.
\begin{eqnarray}
	\Delta_{12}=\frac{7 \theta ^6}{2 \left(\theta ^2+4\right)^6} \left(\theta ^2-32\right) \left(5 \theta ^2-104\right). 
\end{eqnarray}
\begin{center}
	\begin{figure}[!h]
		\centering\includegraphics[totalheight=5cm]{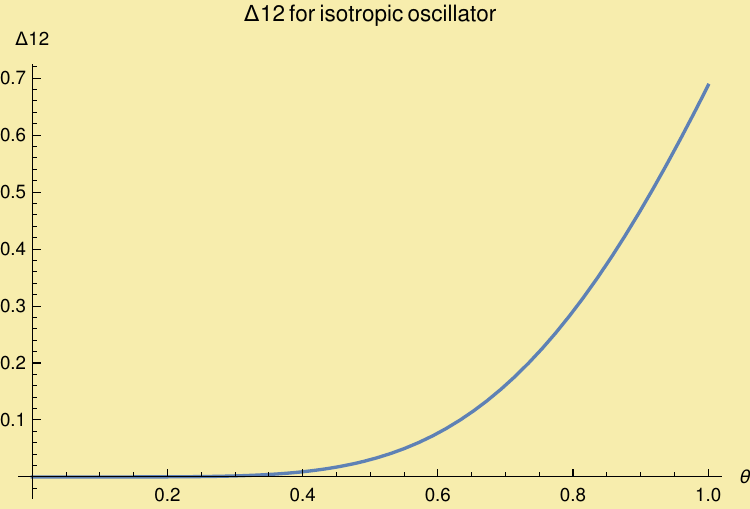}
		\includegraphics[totalheight=5cm]{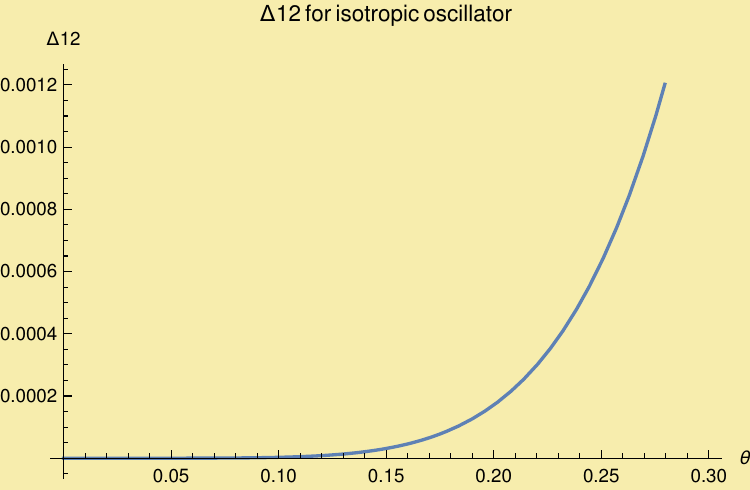}\\
		\caption{\textbf{Isotropic oscillator: Variation of $\Delta_{12}$ with $\theta$. Clearly $\Delta_{12}\ge 0.$ \\
				Parameter values: $m_1=m_2=\tilde{\omega}_1=\tilde{\omega}_2=1$, $\hbar=1$, $\eta=0$. }}\label{delta12theta0to1}
	\end{figure}
\end{center}
For practical purposes, we consider $0\le\theta\le 1$. The variation of $\Delta_{12}$ with respect to $\theta$ may be seen from FIG.~\ref{delta12theta0to1}. In FIG.~\ref{delta12theta0to1}, one can see that $\Delta_{12}$ is always positive for $0\le\theta\le 1$. In other words, we can use  $\vert \Delta_{12}\vert =\Delta_{12}$ in the expression ~\eqref{separabilityc} of separability condition. In particular, for the considered parameter values, $P_s$ reads
\begin{eqnarray}
	P_s^{(isotropic)}(\theta(t))=\frac{1}{32 \left(\theta ^2+4\right)^{12}}(2 \theta ^{24}-19 \theta ^{22}+35592 \theta ^{20}-1992656 \theta ^{18} \nonumber \\ 
	+50248192 \theta ^{16} 
	-622621952 \theta ^{14}+2552965120 \theta ^{12}-8164618240 \theta ^{10} \nonumber \\ 
	+4239065088 \theta ^8  -2346188800 \theta ^6-809500672 \theta ^4+100663296 \theta ^2+33554432). 
\end{eqnarray}

\begin{center}
	\begin{figure}[!h]
		\centering\includegraphics[totalheight=5cm]{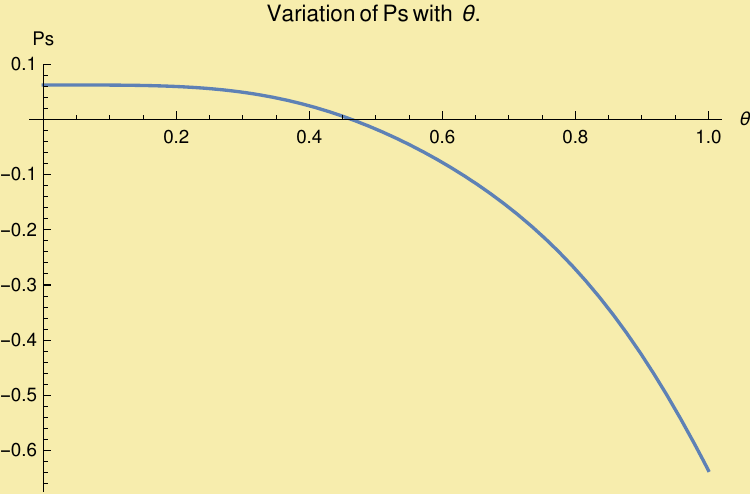}
		\includegraphics[totalheight=5cm]{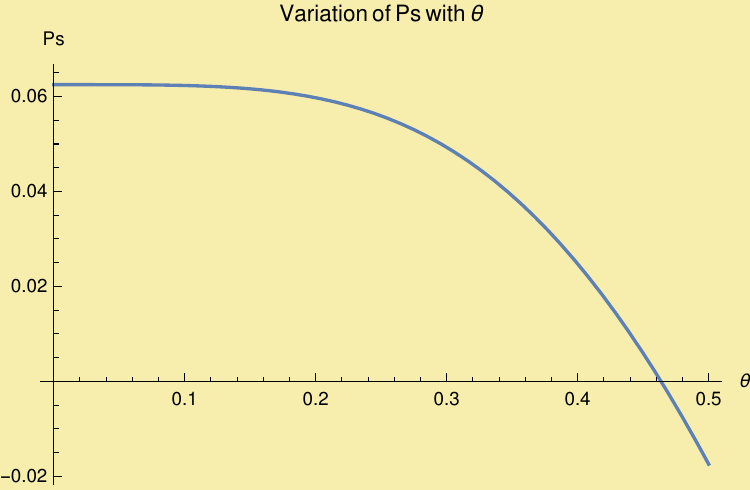}\\
		\caption{\textbf{Isotropic oscillator: Variation of $P_s$ with $\theta$.\\
				$P_s <0 $ corresponds to entangled states, whereas $P_s\ge0$ are separable states. \\
				Parameter values: $m_1=m_2=\tilde{\omega}_1=\tilde{\omega}_2=1$, $\hbar=1$, $\eta=0$. }}\label{Pstheta0to1}
	\end{figure}
\end{center}

\begin{center}
	\begin{figure}[!h]
		\centering\includegraphics[totalheight=5cm]{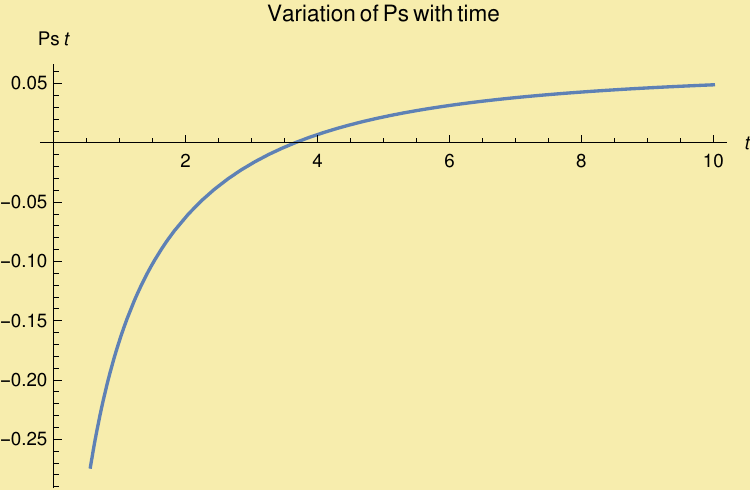}\\
		\caption{\textbf{Isotropic oscillator: Variation of $P_s$ with time ($t$).\\
				$P_s <0 $ corresponds to entangled states, whereas $P_s\ge0$ are separable states. The separability of states varies with time, if the NC-parameter $\theta(t)$ varies with time.\\
				Parameter values: $m_1=m_2=\tilde{\omega}_1=\tilde{\omega}_2=1$, $\hbar=1$, $\eta=0$. }}\label{Pst}
	\end{figure}
\end{center}

To get a glimpse of the separability of oscillator states in NC-space, one can see FIG.~\ref{Pstheta0to1}. We see that by tuning the parameter values $\theta$, one can tune the separability of states.\\
For instance, let us consider that the TD-parameter $\theta(t)$ is monotonically decreasing with time, such that the spatial noncommutativity goes off after a sufficiently large time. In other words, we assume that the dynamics of NC-space is such that it will become a usual commutative space at $t\to\infty$. For a working toy model, let us choose 
\begin{equation}\label{thetatform}
	\theta(t)=1/\sqrt{1 + t}.
\end{equation}
For the choice ~\eqref{thetatform}, we get
\begin{eqnarray}
	P_s(t)= -\frac{1}{32 (4 t+5)^{12}}(-33554432 t^{12}-503316480 t^{11}-2512388096 t^{10} \nonumber \\ -2477260800 t^9 
	+20085276672 t^8+96060977152 t^7+225469802496 t^6 \nonumber \\   +345906574592 t^5 
	+369582065408 t^4 
	+273570431440 t^3 \nonumber \\  
	+133538609768 t^2   +38540796531 t+4968390617),
\end{eqnarray}
which corresponds to the FIG.~\ref{Pst}. From FIG.~\ref{Pst}, we see that initially, the states are entangled, whereas as time elapsed, gradually the states become separable. In particular, the transition between separable states and entangled states occurs at around the close vicinity of  $t \to 3.65086$.
Thus, we see that the entanglement appears due to the specific NC-parameter values. In the next section, we have outlined a toy model to get a more physical glimpse.  
\section{Toy Model: Anisotropic oscillator with only spatial noncommutativity}
Once again we consider the NC-space commutation relations ~\eqref{toyncalgebra}. For quantitative discussion, we set the following parameter values for the anisotropic oscillator.
\begin{equation}\label{parametervalues}
\hbar\to 1, \; m_1=1,m_2=4, \tilde{\omega}_1=\tilde{\omega}_2=1.
\end{equation}
For the present study, let us assume that the TD-parameter $\theta(t)$ is monotonically decreasing with time, such that the spatial noncommutativity goes off after a sufficiently large time. In other words, we assume that the dynamics of NC-space is such that it will become a usual commutative space at $t\to\infty$. For convenience, again we consider ~\eqref{thetatform}. 
Using the parameters ~\eqref{parametervalues} and ~\eqref{thetatform}, we get the following explicit form of $Ps$ (as mentioned in ~\eqref{separabilityc}) for the state $\vert 0,0;\beta_1,\beta_2\rangle$.
\begin{eqnarray}
Ps=\frac{1}{16(2+t)^{12}} [ t^{12} + 3\times 2^3 t^{11} - 2^3 (14431 + 2^{11} t_7 ) t^{10} - 2^5 ( 78689 + 10264 t_7 ) t^9 \nonumber \\
+ 2^4 (204010327 +  51198584 t_7 ) t^8 + 2^5 ( 1267439685 + 272489392 t_7 ) t^7 \nonumber \\
+ 2^5 ( 6717167061 + 1248828188 t_7 ) t^6 + 2^8 ( 2486103583 + 402820658 t_7 ) t^5 \nonumber \\
+ 2^7 ( 9004286997 + 1281090628 t_7 ) t^4 + 2^9 ( 2555083273 + 321598952 t_7 ) t^3 \nonumber \\
+ 2^9 (1776825135 + 199309604 t_7) t^2 + 2^{13} ( 43306567 + 4360181 t_7) t \nonumber \\
+ 2^{11} ( 29038819 + 2642044 t_7 )],\; \mbox{with} \; t_7=\frac{\sqrt{7(15+8t)}}{1+t}.
\end{eqnarray}
If $Ps \ge0$, then the NC-space co-ordinate degrees of freedoms are separable, otherwise they are entangled. One can verify that
\begin{equation}
	\lim\limits_{t\to \infty} Ps =1/16,
\end{equation}
which confirms that NC-space is gradually transformed into usual commutative space through the dynamics of space. 
\begin{center}
	\begin{figure}[!h]
		\centering\includegraphics[totalheight=6cm]{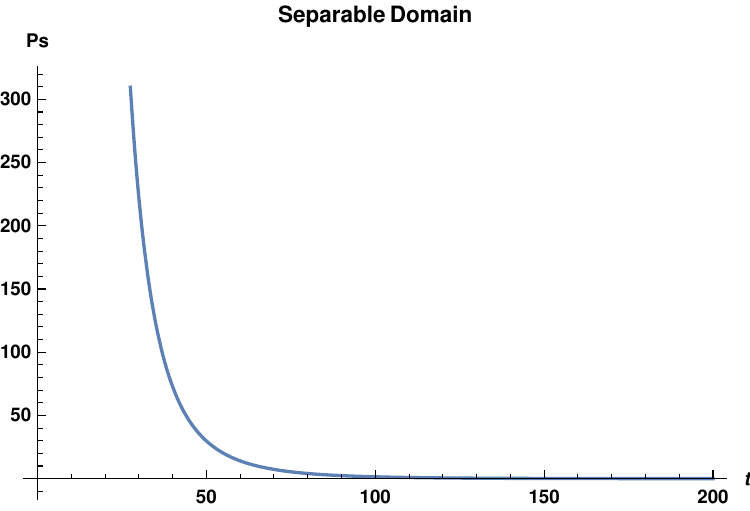}\\
		\caption{\textbf{$Ps\ge0 \implies$ separable state  }}\label{sampleFig1}
	\end{figure}
\end{center}
\begin{center}
\begin{figure}[!h]
	\centering\includegraphics[totalheight=6cm]{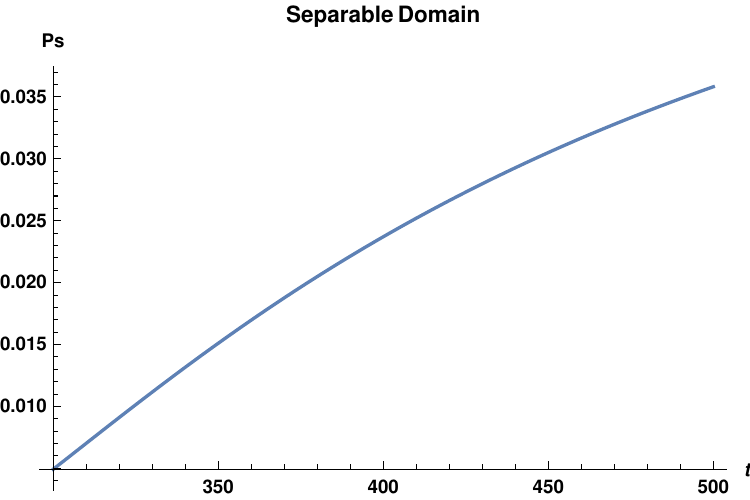}\\
	\caption{\textbf{$Ps\ge0 \implies$ separable state  }}\label{sampleFig2}
\end{figure}
\end{center}
\begin{center}
		\begin{figure}[!h]
			\centering\includegraphics[totalheight=6cm]{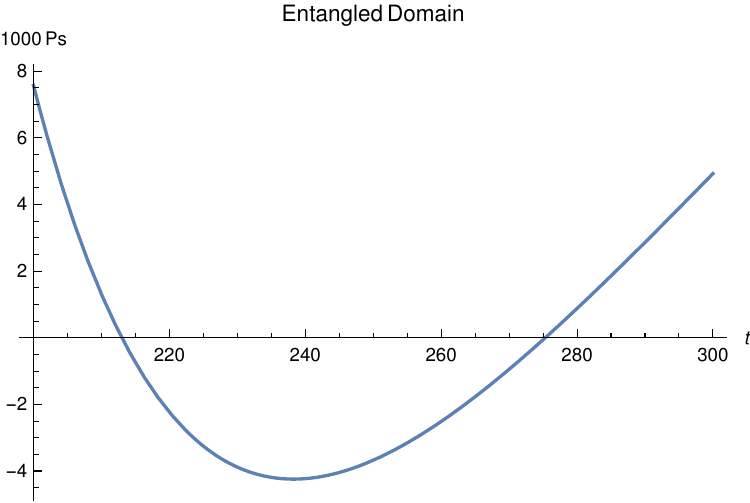}\\
			\caption{\textbf{$Ps<0 \implies$ Entangled state  } ($1000Ps$ is plotted for better understanding of figure, since the negative values are too small for graphical representation.)
			}\label{sampleFig3}
		\end{figure}
	\end{center}
Since the maximum value of $Ps\sim 10^6$, whereas the minimum value of $Ps \sim -10^{-3}$, it is difficult to visualize the nature of $Ps$ in a single plot. We have divided the three regions for $Ps$ in which it is positive or negative and show the transition between positive and negative values. FIG.~\ref{sampleFig1} and FIG.~\ref{sampleFig2} represent the separable domain, whereas FIG.~\ref{sampleFig3} represents entangled domain. Interestingly, there is a two-way transition- after the time at around $t \to 213.001$ separable states become entangled, whence it becomes separable again approximately after $t\to 275.331$. 
\section{Conclusions}
 Due to the nonsymplectic connection between the covariance matrices of noncommutative space (NCS) and the commutative space,  the generalized Peres-Horodecki criterion (Simon's condition) for the separability of bipartite Gaussian state is not immediate in NCS, despite their structural similarity. We have outlined a detailed study of the mechanism of indirect utilization of Simon's condition for NCS through the Bopp shift for both the position-position and momentum-momentum non-commutativity.  
 \\
 We have shown that the equivalent system in commutative space can be transformed into canonical form maintaining the symplectic (group $Sp(4,\mathbb{R})$) structure intact.
 The lewis-Riesenfeld phase-space invariant method is utilized to construct the exact solution (state of the system) of the time-dependent (TD) Schr\"{o}dinger equation for the concerned TD-Hamiltonian with bilinear and linear field modes interactions in commutative space.  The exact form of solutions of the TD system enabled us to construct the exact form of the required covariance matrix.  It turns out that the covariance matrix looks the same as that of the ordinary oscillator in commutative space, with the only difference being in the matrix elements, which are dependent on NCS parameters.
 \\
It turns out that the existence of the deformation parameters ($\theta$ and $\eta$) in NC-space determines the separability criteria of the bipartite Gaussian state of oscillators. In particular, for an isotropic oscillator, the separability of the bipartite state depends on the TD-NC parameter values. On the other hand, anisotropic parameter values may cease the separability of the bipartite state in NC-space. This observation opens up the possibility of engineering the entanglement by tuning the parameter of the system. In particular, since oscillators in NCS are structurally similar to that of the dynamics of a charged particle inside a magnetic field (e.g., Maxwell Chern-Simons model in long wavelength limit in $2$ spatial dimension), the time evolution of NCS parameters is equivalent to a time-dependent external magnetic field, which can be controlled in a laboratory environment, with present-day experimental capacity. For a simple illustration of the experimental possibility, we have outlined  simple toy models by considering only spatial noncommutativity and anisotropy in effective mass.  The transition point of the switching between separability and entanglement are illustrated graphically.
 \section{Acknowledgement} 
Pinaki Patra is grateful to the Science and Engineering Research Board (under the Department of Science and Technology, Government of India) for financial support through a project grant (File Number: EEQ/2023/000784).\\
 We are grateful to the anonymous referee(s) for valuable comments and suggestions, which made the manuscript in its present form.
 \section{Author Declarations}
 The authors have no conflicts to disclose.
 \section{Availability of data}
 Data sharing is not applicable to this article as no new data were created or analyzed in this study.

\renewcommand{\theequation}{A.\arabic{equation}}

\setcounter{equation}{0}
\begin{appendices}
	\section{Eigenvalues  of $\hat{\Omega}$}
  The characteristic polynomial  
	($
	p(\lambda)= \det (\lambda \mathbb{\hat{I}}- \hat{\Omega}) ,
	$)
	of  $\hat{\Omega}$ is given by
	\begin{equation}
		p(\lambda)= \lambda^4+ b\lambda^2 +c .
	\end{equation}
	Where
		\begin{equation}
		c = \omega_x^2 \omega_y^2 ,\;\;
		b = \alpha_0 \omega_x^2 + \frac{1}{\alpha_0}\omega_y^2 + 4\nu_1\nu_2 (\sqrt{\alpha_0}+\frac{1}{\sqrt{\alpha_0}})^2 ,
	\end{equation}
	with
	\begin{equation}
		\omega_x^2 = 4\nu_1\nu_2 (\frac{\mu_1\omega_1^2}{4 \mu_2\nu_2^2} -1) ,\;
		\omega_y^2 = 4\nu_1\nu_2 (\frac{\mu_2\omega_2^2}{4 \mu_1\nu_1^2} -1), \;
		\alpha_0 = \frac{\mu_2\nu_2}{\mu_1\nu_1}.
	\end{equation}
	Using the explicit forms of $\mu_i,\nu_i,\omega_i,\; i=1,2$, one can show that 
	$
	c \ge 0,\; b\ge 0
	$ as follows.\\
	First we observe that $b$ and $c$ can be written as
	\begin{equation}
		b = \omega_1^2 + \omega_2^2 + 6\nu_1\nu_2 ,\;
		c = \omega_1^2  \omega_2^2 + 16 \nu_1^2 \nu_2^2 -4\nu_1^2 \omega_1^2 \mu_1/\mu_2 - 4\nu_2^2 \omega_2^2 \mu_2/\mu_1 .
	\end{equation}
	$b>0$ for nonzero positive values of $\omega_1, \omega_2, \nu_1, \nu_2$. On the other hand,
	 $c$ can be factorized as
	\begin{equation}
		c= 16\nu_1^2\nu_2^2 \left( \frac{\mu_1\omega_1^2}{4\mu_2\nu_2^2}-1 \right) \left( \frac{\mu_2\omega_2^2}{4\mu_1\nu_1^2}-1 \right).
	\end{equation}
	If we use the explicit forms of $\omega_1, \omega_2, \nu_1, \nu_2$, we have
	\begin{eqnarray}
		\frac{\mu_1\omega_1^2}{4\mu_2\nu_2^2} &=& \frac{4m_2^2\hbar^2}{(\eta+ m_1m_2\theta \tilde{\omega}_1^2)^2} \left(m_1\tilde{\omega}_1^2+ \frac{\eta^2}{4\hbar^2 m_2}\right) \left( \frac{1}{m_2} + \frac{1}{4\hbar^2}m_1\tilde{\omega}_1^2 \theta^2 \right) \nonumber \\
		&=& \left(1+ \frac{4\hbar^2}{\eta^2}m_1m_2 \tilde{\omega}_1^2 \right)  \left(1+ \frac{\theta^2}{4\hbar^2}m_1m_2 \tilde{\omega}_1^2 \right) \left(1+ \frac{\theta}{\eta}m_1m_2 \tilde{\omega}_1^2 \right)^{-2} \nonumber \\
		&=& \frac{1}{\left(1+ \frac{\theta}{\eta}m_1m_2 \tilde{\omega}_1^2 \right)^{2}}\left( \left(1+ \frac{\theta}{\eta}m_1m_2 \tilde{\omega}_1^2 \right)^{2} + \frac{2\theta}{\eta}m_1m_2\tilde{\omega}_1^2 (\frac{2\hbar^2}{\eta\theta}+ \frac{\eta\theta}{8\hbar^2}-1)\right) \nonumber \\
		&=& 1+ \frac{2\theta m_1m_2 \tilde{\omega}_1^2}{\eta \left(1+ \frac{\theta}{\eta}m_1m_2 \tilde{\omega}_1^2 \right)^{2}} \left( \frac{\sqrt{2}\hbar}{\sqrt{\eta\theta}} - \frac{\sqrt{\eta\theta}}{2\sqrt{2}\hbar} \right)^2  
		\ge 1 .
	\end{eqnarray}
	Similarly, we see that
	\begin{equation}
		\frac{\mu_2\omega_2^2}{4\mu_1\nu_1^2} = 1+ \frac{2\theta m_1m_2 \tilde{\omega}_2^2}{\eta \left(1+ \frac{\theta}{\eta}m_1m_2 \tilde{\omega}_2^2 \right)^{2}} \left( \frac{\sqrt{2}\hbar}{\sqrt{\eta\theta}} - \frac{\sqrt{\eta\theta}}{2\sqrt{2}\hbar} \right)^2  \ge 1.
	\end{equation}
	Therefore, one can conclude
	$
		c\ge 0.
	$
	
	Moreover, the discriminant of the characteristic polynomial 
	\begin{equation}
		\Delta 
		= \left(\alpha_0 \omega_x^2 -\frac{1}{\alpha_0}\omega_y^2\right)^2 + 8\nu_1\nu_2 (1+ \alpha_0)^2 (\omega_x^2 +\omega_y^2) + \frac{4\nu_1^2 \nu_2^2}{\alpha_0^2} (1+\alpha_0)^4 \ge 0.
	\end{equation}
	One can easily verified that
	\begin{equation}\label{lambdasquare}
		\lambda^2 = \frac{1}{2}(-b\pm \sqrt{\Delta}) \le 0.
	\end{equation}
	From ~\eqref{lambdasquare}, we can conclude that $\hat{\Omega}$ has four distinct imaginary eigen-values
	\begin{eqnarray}
		\lambda=\left\{ -i\lambda_1,i\lambda_1, -i\lambda_2,i\lambda_2 \right\}.
	\end{eqnarray}
	Where
	\begin{equation}
		\lambda_1 = \frac{1}{\sqrt{2}}\sqrt{b+\sqrt{\Delta}} ,\;
		\lambda_2 = \frac{1}{\sqrt{2}}\sqrt{b-\sqrt{\Delta}} .
	\end{equation}
\end{appendices}

\end{document}